\documentclass[12pt,oneside,reqno]{amsart} 
\usepackage{amssymb}
\usepackage[usenames, dvipsnames]{color}
\usepackage{graphicx}
\usepackage{float}
\usepackage{wrapfig}
\restylefloat{figure}
\usepackage{setspace}

\usepackage{amsmath,amsfonts,epsfig,graphicx,float,gensymb,esvect,nicefrac}
\usepackage{array}
\usepackage{comment}
\newcolumntype{P}[1]{>{\centering\arraybackslash}p{#1}}
\newcolumntype{M}[1]{>{\centering\arraybackslash}m{#1}}

\usepackage[super,square,numbers, sort&compress]{natbib}
\begin{document}

%%%%%%%%%%%%%%%%%%%%%%%%%%%%%%%%%%%%%%%%%%%%%%%%%%%%%%%%%%%%%%%%%%%%%

%\noindent{DOI: 10.1002/ ((please add manuscript number))}
%
%\noindent{\textbf{Article type: Full Paper}}\\

%%%%%%%%%%%%%%%%%%%%%%%%%%%%%%%%%%%%%%%%%%%%%%%%%%%%%%%%%%%%%%%%%%%%%

%%%%%%%%%%%%%%%%%%%%%%%%%%%%%%%%%%%%%%%%%%%%%%%%%%%%%%%%%%%%%%%%%%%%%

\noindent{\textbf{Longitudinal Eigenvibration of Multilayer Colloidal Crystals and the Effect of Nanoscale Contact Bridges}} \\

%%%%%%%%%%%%%%%%%%%%%%%%%%%%%%%%%%%%%%%%%%%%%%%%%%%%%%%%%%%%%%%%%%%%%

\noindent Maroun Abi Ghanem$^{1}$, Amey Khanolkar$^{1}$, Samuel P. Wallen$^{2}$, Mary Helwig$^{3}$, Morgan Hiraiwa$^{3}$, Alexei A. Maznev$^{4}$, Nicolas Vogel$^{5}$ and Nicholas Boechler$^{1}$\\

$^{1}$ Department of Mechanical and Aerospace Engineering, University of California, San Diego, La Jolla, CA 92093 USA

$^{2}$ Applied Research Laboratories, The University of Texas at Austin, Austin, TX, 78758 USA

$^{3}$ Department of Mechanical Engineering, University of Washington, Seattle, WA, 98195 USA

$^{4}$ Department of Chemistry, Massachusetts Institute of Technology, Cambridge, MA 02139 USA

$^{5}$ Institute of Particle Technology, Friedrich-Alexander University, Erlangen-Nurnberg, Cauerstrasse 4, 91048 Erlangen, Germany

%%%%%%%%%%%%%%%%%%%%%%%%%%%%%%%%%%%%%%%%%%%%%%%%%%%%%%%%%%%%%%%%%%%%%

Keywords: colloidal crystals, eigenvibrations, contact dynamics, adhesion, granular crystals\\

%%%%%%%%%%%%%%%%%%%%%%%%%%%%%%%%%%%%%%%%%%%%%%%%%%%%%%%%%%%%%%%%%%%%%%%%%%%%%

%===================
%== ABSTRACT ===
%===================

Longitudinal contact-based vibrations of colloidal crystals with a controlled layer thickness are studied. These crystals consist of $390$~nm diameter polystyrene spheres arranged into close packed, ordered lattices with a thickness of one to twelve layers. Using laser ultrasonics, eigenmodes of the crystals that have out-of-plane motion are excited. The particle-substrate and effective interlayer contact stiffnesses in the colloidal crystals are extracted using a discrete, coupled oscillator model. Extracted stiffnesses are correlated with scanning electron microscope images of the contacts and atomic force microscope characterization of the substrate surface topography after removal of the spheres. Solid bridges of nanometric thickness are found to drastically alter the stiffness of the contacts, and their presence is found to be dependent on the self-assembly process. Measurements of the eigenmode quality factors suggest that energy leakage into the substrate plays a role for low frequency modes but is overcome by disorder- or material-induced losses at higher frequencies. These findings help further the understanding of the contact mechanics, and the effects of disorder in three-dimensional micro- and nano-particulate systems, and open new avenues to engineer new types of micro- and nanostructured materials with wave tailoring functionalities via control of the adhesive contact properties.\\

%%%%%%%%%%%%%%%%%%%%%%%%%%%%%%%%%%%%%%%%%%%%%%%%%%%%%%%%%%%%%%%%%%%%% %%%%%%%%

%===================
%== Introduction ===
%===================

\textbf{1.	Introduction}\\

Colloidal self-assembly has emerged as a ``bottom-up" approach to manufacturing nanostructured materials with controlled hierarchical architectures, and has enabled their fabrication inexpensively and on a large scale \cite{VogelReview}. Primarily investigated for their unprecedented features in areas like photonics \cite{Freymann,AkimovPRL2008}, plasmonics \cite{Klinkova}, and biosensing \cite{Alivisatos, Sepulveda}, self-assembled colloidal crystals, which consist of ordered, close-packed lattices of micro- to nanoscale spherical particles \cite{JiangSelfAssembly}, have also shown promise in the context of their vibrational dynamics or ``phononic'' properties \cite{NedThomasReview2,FytasNatureMaterials}. \\

In a colloidal crystal, particles may be dispersed in a liquid or in a solid matrix, or the host medium may be absent (e.g. a ``dry'' colloidal crystal) \cite{FootnoteColloid,AizenbergReview2016,LopezReview2016}. Macroscale analogues of dry colloidal crystals, often referred to as ``granular crystals,'' have also been a topic of significant interest due to their unique dynamics \cite{NesterenkoBook, GranularCrystalReviewChapter, PhysicsTodayGranular}. In granular crystals, spheres arranged in close packed arrays can be thought to move similar to rigid bodies and interact via small contact regions (compared to the particle size) that elastically deform and act as massless springs. Such contact-based dynamics have underlied the large interest in granular crystals, as the interplay of the contact nonlinearity, typically modeled by the Hertzian contact model \cite{HertzPaper}, coupled with dispersion induced by mechanisms such as structural periodicity \cite{GranularCrystalReviewChapter} and local resonances \cite{BonanomiLocalResonance2015, TournatRotation, TournatMagnetoGranular}, have been shown to enable new acoustic wave tailoring strategies \cite{GranularCrystalReviewChapter, PhysicsTodayGranular}. However, until recently, studies concerning the contact-based dynamics of granular crystals have been restricted to systems composed of millimeter to centimeter-sized spheres. Acoustic studies of colloidal crystals, on the other hand, mostly focused on colloids in a host medium \cite{Fytas2012,FytasNatureMaterials,Still2008}. Until recently, the contact-based dynamics of dry colloidal crystals or ``micro- to nanoscale granular crystals'', wherein the Hertzian contacts are preserved and play a major role, remained an unexplored territory. \\

The contact-based dynamics of dry, two-dimensional (2D) colloidal crystal monolayers have been examined in several recent studies \cite{Boechler_PRL, EliasonMicrosphereStrip, HiraiwaThreeResonance,Wallen2015,VegaFlickExp,VegaFlickSpheroidalTheory}. In these works, it was shown that adhesive van der Waals forces \cite{Israelachvili} statically compress the interparticle and particle-substrate contacts, and effectively linearized the crystal dynamics for small displacements (relative to the static deformation). This compression resulted in the formation of multiple contact resonances of the crystal, including modes with both translational as well as coupled translational-rotational motion  \cite{Wallen2015,HiraiwaThreeResonance,VegaFlickExp,VegaFlickSpheroidalTheory}. In contrast, the contact-based dynamics of three-dimensional (3D) colloidal crystals remain largely unexplored. Using a Brillouin Light Scattering spectroscopic technique, a previous study exploring 3D colloidal crystal dynamics revealed a single resonant peak that was attributed to a band of contact-based modes, however, the individual eigenmodes were not resolved \cite{Fytas2012}. Laser ultrasonic techniques have also been used to measure the transmission of hypersonic acoustic waves traveling through dry 3D colloidal crystals \cite{AkimovPRL2008,JACS2005}. In contrast to the study presented herein, the vibrational modes observed in Reference \cite{JACS2005} were eigenmodes of isolated spheres (with frequencies significantly higher than the contact-based modes concerned in this work). A presence of an overdamped ``low-frequency continuum mode'' that was attributed to shear waves was also suggested in Reference \cite{JACS2005}, however its frequency was not identified. Similarly, the measurements presented in Reference \cite{AkimovPRL2008} provided limited information concerning the acoustic modes of the crystal, no characterization of the interparticle contact stiffnesses, and, due to the sintering of the crystal, we expect limited dynamic deformation of the particle contacts was involved. Several studies have also explored the contact-mediated acoustics of 3D colloidal films, however these were either disordered \cite{Ayouch2012,Ruello2017}, or the spheres were separated by large ligands  \cite{Lisiecki2013, Gusev2015PRB, Poyser2015} or polymer tethers \cite{Fytas2014} that approach the particle size and result in non-Hertzian contact mechanics. \\

In this work, we study the longitudinal contact-based vibration of self-assembled, dry, 3D colloidal crystals with long range order of varied thicknesses, and measure multiple, discrete eigenmodes of the crystal. In contrast to prior studies, using our laser ultrasonic technique, we are able to access the low-frequency, contact-based modes of the 3D crystals, and our spheres are not functionalized with large connective ligands that can cause deviation from the Hertzian mechanics of elastically deforming spheres in contact. We correlate the resulting contact stiffnesses, extracted by way of a coupled-oscillator model, with scanning electron microscopy (SEM) and atomic force microscopy (AFM) observation of the contacts, and find that nanometric bridges surrounding the contacts, which are much smaller than the particle size, can more than triple the contact stiffness. In all cases, we find higher contact stiffnesses for our self-assembled crystals than those predicted using adhesive elastic contact models \cite{DMT1,Mindlin}. We find also that the bridge sizes and contact stiffness vary with sample fabrication method, and that the contact stiffness can decrease with increasing numbers of crystal layers. By studying the quality factors of the measured eigenmodes, and comparing them with an analytical model that estimates acoustic energy radiation into the substrate based on the estimated impedance mismatch between the colloidal crystal and the substrate, we gain insight into energy loss mechanisms in our system. From this comparison, we suggest that energy leakage plays a large role for low frequency modes but is surpassed by disorder- or material-induced losses as the modal frequency increases. Improved understanding of the dynamics of self-assembled, dry, 3D colloidal crystals will help enable the design of new, microstructured materials with ultrasonic wave tailoring capabilities that leverage contact-based nonlinearities analogous to those present in macroscale granular crystals\cite{NesterenkoBook, GranularCrystalReviewChapter, PhysicsTodayGranular}. We expect that potential applications for such materials may range from signal processing to impact mitigation and energetic material design.   \\

\textbf{2.	Results and Discussion}\\
 %%=========================
%% ==Results and Discussion===
%%=========================

\begin{figure}
 \includegraphics[width=10.5cm]{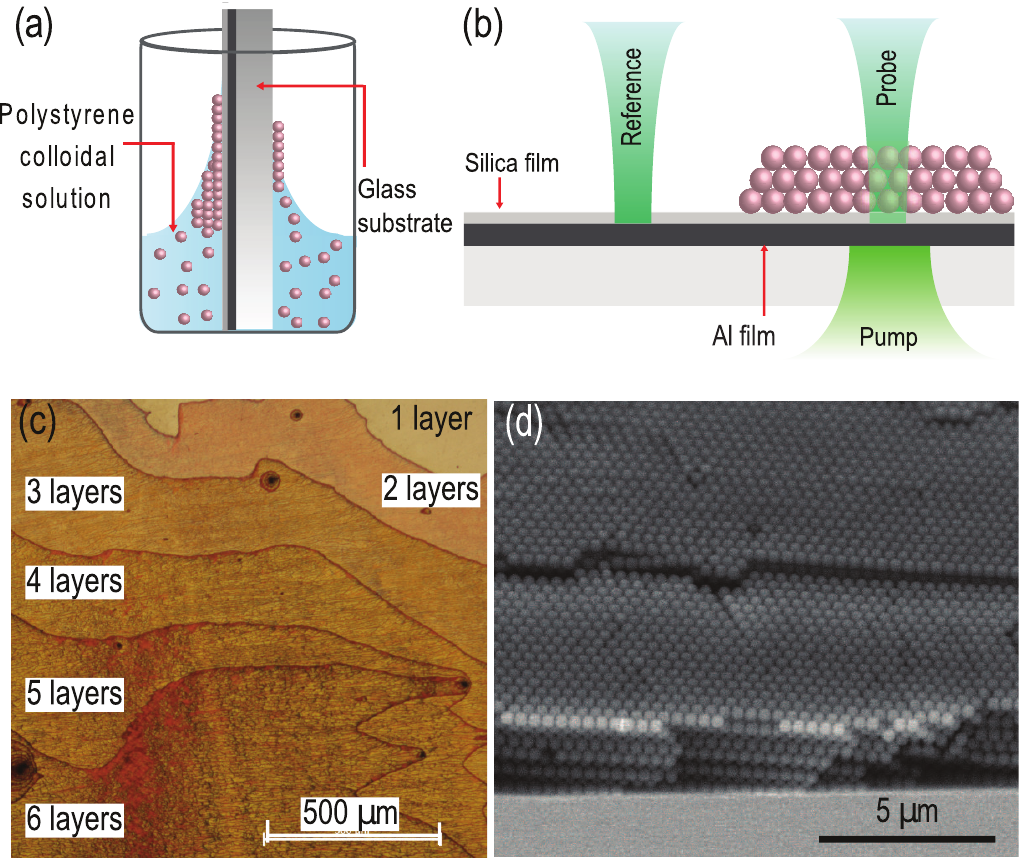}
 \caption{(a) Schematic of the multilayer convective self-assembly technique. (b) Illustration of the laser ultrasonic technique used to excite and measure eigenmodes of the colloidal crystal. (c) Optical microscope image showing multiple regions of the colloidal crystal with different layer thicknesses. (d) Representative SEM image of the colloidal crystal. }
 \label{fgr:fig1}
\end{figure}

Our colloidal crystals consist of close-packed, ordered arrangements of polystyrene spheres of diameter $D$ = $390$~nm, which are assembled on an aluminum-coated glass microscope slide, and are fabricated by vertical convective self-assembly \cite{JiangSelfAssembly, MeijerSelfAssembly}, as shown in \textbf{Figure~\ref{fgr:fig1}}(a). Regions with uniform layer thickness can be easily identified under a microscope from their different coloration, as is illustrated in Figure~\ref{fgr:fig1}(c). In the most densely packed case, this geometry represents perfectly registered stacks of hexagonally close-packed (HCP) or face-centered-cubic (FCC) colloidal monolayers. A representative SEM image of our colloidal crystal can be seen in Figure~\ref{fgr:fig1}(d). The distinct structural coloration enables us to identify regions with defined layer thickness and thus study, in detail, the thickness-dependent contact-based acoustic properties of the crystals. To characterize the longitudinal acoustic response of the colloidal crystals, we use a laser ultrasonic technique that is illustrated in Figure~\ref{fgr:fig1}(b). Absorption of the pump pulse energy by the aluminum film induces a rapid thermal expansion that excites mechanical vibration of the crystal and launches acoustic waves in the substrate. In the colloidal crystal, longitudinal vibrations with out of plane motion are predominantly excited, as the excitation of transverse vibrations are hindered by symmetry constraints. These vibrations are detected with a phase-mask-based interferometer \cite{GlorieuxInterferometer2004}, which is preferentially sensitive to out-of-plane displacements \cite{HiraiwaThreeResonance}. As a result of the semi-transparency of the colloidal crystal, we expect that the probe signal contains contributions due to displacement of the surface of the colloidal crystal, scattering and refractive index changes within the colloidal crystal, and displacement of the aluminum-silica interface.\\

\begin{figure}
 \includegraphics[width = 11.5cm]{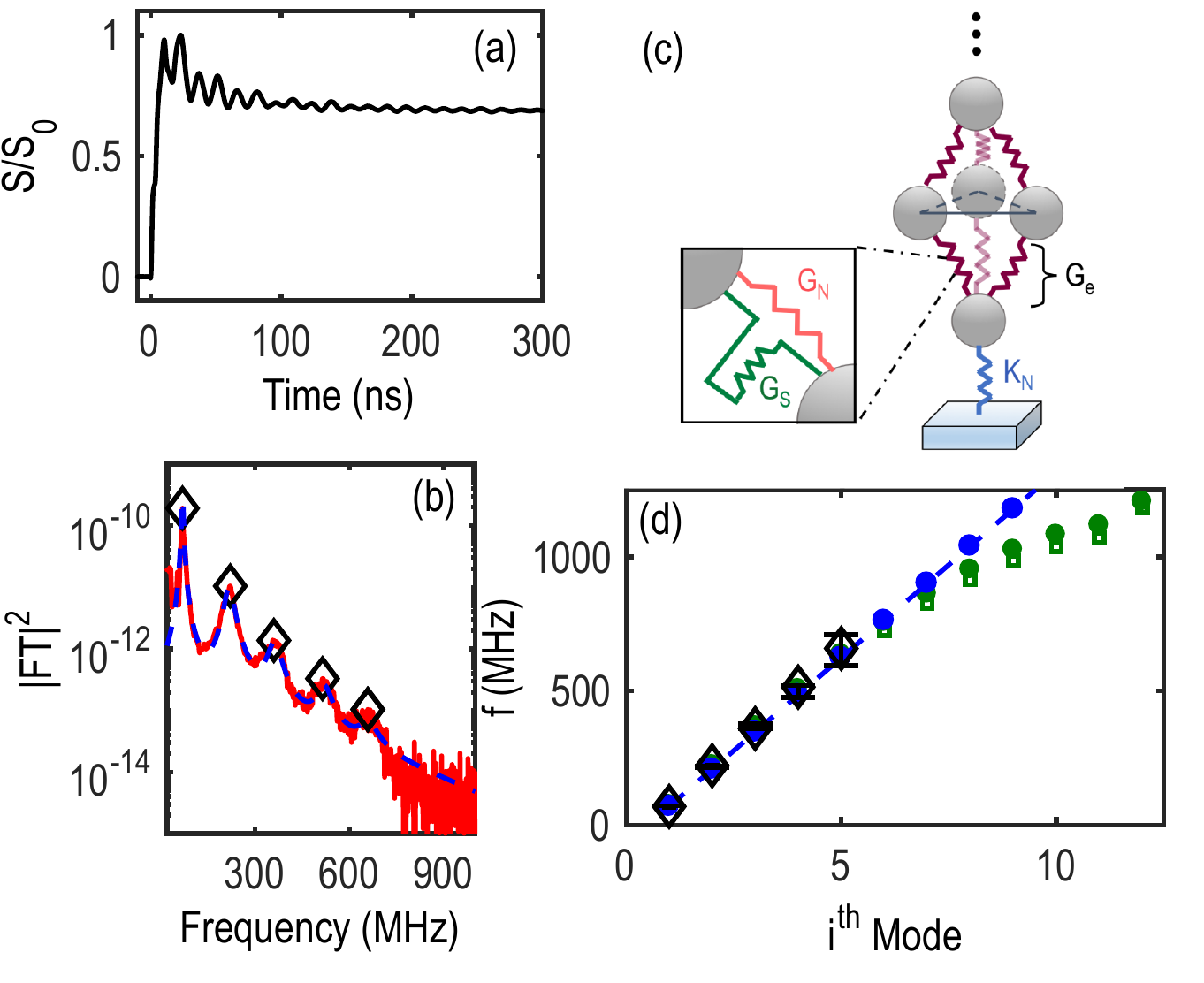}
 \caption{(a) Time domain signal corresponding to the out-of-plane eigenvibrations of a 12-layer-thick colloidal crystal. The signal amplitude $S$ is normalized to its maximum amplitude $S_{0}$. (b) The solid red line denotes the power spectrum of the time-derivative of the signal in (a), and the dashed blue line denotes the sum of five Lorentzians fitted to the measured spectrum. (c) Schematic of the quasi-one-dimensional coupled oscillator model. (d) Modal frequencies as a function of mode number. Black diamond markers are the modes identified in (b) denoted by the same marker type. The blue circle markers represent the calculated modal frequencies for a fixed-free continuum film adhered to a rigid substrate, where the first mode is matched to the measured fundamental mode. The blue dashed line is a visual guide to the blue circle markers, and represents a wave speed of 1060 m/s. The green markers represent the calculated modal frequencies of the coupled oscillator system using a particle-substrate stiffness obtained via a monolayer region of the same sample measured in (a,b) and an interlayer contact stiffness fitted to the fundamental measured mode (open square markers), and to all five measured modes (filled circle markers). The error bar half-widths in the measured spectral peaks denote the maximum shift in the position of the peaks when the power spectrum time window is adjusted by up to $4$~ns.}
 \label{fgr:fig2}
\end{figure}

\textbf{Figure~\ref{fgr:fig2}}(a) shows a signal acquired on a twelve-layer-thick region of a colloidal crystal sample, using the optical characterization method shown in Figure~\ref{fgr:fig1}(b). A sharp initial rise due to the arrival of the pump pulse at the sample followed by a slow decay associated with thermal diffusion is observed. The periodic oscillations in the signal represent the out-of-plane longitudinal vibrations of the colloidal crystal. The power spectrum (amplitude squared of the Fourier transform spectrum) of the time-derivative of the signal in Figure~\ref{fgr:fig2}(a) is shown in Figure~\ref{fgr:fig2}(b), wherein five spectral peaks can be distinguished. The fundamental peak has a frequency $f_1=70$~MHz. The intensities of the subsequent peaks decrease with increasing frequency, such that the intensity of the fifth peak is over three orders of magnitude lower than that of the fundamental peak. We suggest that this is due, in part, to the step-like nature of the excitation, which more efficiently excites modes at lower frequencies, and the increased susceptibility of shorter-wavelength, higher-frequency modes to disorder-induced scattering \cite{TournatReview}. Each of the modes identified in the spectrum shown in Figure~\ref{fgr:fig2}(b) are denoted by open diamond markers, and are plotted as function of mode number in Figure~\ref{fgr:fig2}(d) using the same markers. We compare the measured frequencies to the modal frequencies of a continuous film with boundaries that are fixed on one side and free on the other, given by $f_{i} =  \frac{(2i - 1)c}{4H}$, where $i$ is the mode number, and $c$ and $H$ are the longitudinal sound speed and thickness, respectively of the film \cite{Poyser2015}. In our sample, the substrate and the crystal are mechanically coupled, with the substrate not strictly rigid. Hence, we expect much of the measured vibrations to be confined to the colloidal crystal, due to the large impedance mismatch between the crystal and the substrate. The modal frequencies of the crystal fixed to the substrate on one side are plotted using blue filled circle markers in Figure~\ref{fgr:fig2}(d), where the sound speed of the crystal is chosen such that the frequency of the fundamental mode of the continuous film matches the experimentally measured fundamental mode. From Figure~\ref{fgr:fig2}(d), we see that the five measured modes approximately follow the linear trend of the calculated fixed-free film modes as the mode number increases. Similar observations have been made in previous studies on nanocrystal superlattices interacting via ligands that have shown equally-spaced modes \cite{Poyser2015}. \\

While the observed modes are well described by the linear relationship corresponding to the continuous film model, we develop a quasi-one-dimensional coupled oscillator model to describe our system and analyze the contact stiffnesses, in which we treat the spheres as rigid bodies connected by massless springs. Similar springs-in-series models (without the lumped masses) have been previously used to study the quasi-static nanoindentation of multilayer colloidal crystals composed of hollow silica nanospheres \cite{Retsch2012}. We implement a discrete element model because the measured frequencies are lower than the lowest intrinsic spheroidal mode of the particles ($2.62$~GHz) \cite{Sato1962}, and the discrete model enables us to separately describe particle-substrate and interlayer contact stiffnesses. A schematic of our quasi-one-dimensional coupled oscillator model is illustrated in Figure~\ref{fgr:fig2}(c). The particle-substrate contact between the first layer and the substrate is represented by an axial contact spring of stiffness, $K_{N}$, while the interparticle normal and shear contacts involved in subsequent layers are represented by axial and transverse springs of stiffness $G_{N}$ and $G_{S}$, respectively. The normal and shear stiffnesses are modeled using linearized Hertzian \cite{HertzPaper} and Hertz-Mindlin contact models \cite{Mindlin}. We define an effective interlayer contact stiffness $G_{e}$ oriented along the the out-of-plane direction, which incorporates the contributions from the three pairs of interparticle normal and shear contact springs, and can be expressed as 
\begin{equation} 
G_{e} = G_{N} (2 + \nu^*),
\label{G_eff}
\end{equation}
where for statically compressed Hertzian and Hertz-Mindlin contacts, $\nu^* = \frac{G_{S}}{G_{N}} = 2 \frac{1-\nu_{P}}{2-\nu_{P}}$ is the ratio of the interparticle shear and normal contact stiffnesses, and $\nu_{P}$ is the Poisson's ratio of the polystyrene spheres (see Supporting information and Reference \cite{Mindlin,Merkel}). Using \textbf{Equation~\ref{G_eff}}, the colloidal crystal can be simplified to a quasi one-dimensional coupled oscillator system, where each of the oscillators have mass per unit area $M$, and are connected to one another by springs of stiffness per unit area $G$. Here, $M = m/A_{p}$ and $G = G_{e}/A_{p}$, where $m$ is the mass of one of the spheres and $A_{p} = \sqrt{3} D^2/2$ is the area of the primitive unit cell in the HCP lattice. Although the crystal may also be in the FCC configuration, we assume HCP packing for the purpose of this article. The equivalent one-dimensional coupled oscillator system reduces to a chain of spheres of mass $m$ connected to one another by springs of stiffness $G_{e}$, and connected to the substrate by a spring of stiffness $K_{N}$. The mass of the sphere is calculated using the density $\rho$ = $1060$~kg/m$^{3}$ of polystyrene \cite{KhanolkarLambWaves2015}. Since our measurements are sensitive to the out-of-plane direction only and we excite primarily longitudinal plane waves, we neglect adhesive contact forces in the transverse direction between neighboring spheres within the same layer.\\

We first apply our quasi-one-dimensional coupled oscillator model to the spectrum shown in Figure~\ref{fgr:fig2}(b). Using a single oscillator model, we first determine the particle-substrate contact stiffness $K_{N}$ by measuring the out-of-plane contact resonance frequency $f_{1,m}$ for a monolayer region of the same sample and using the relation $K_{N}$ = $(2\pi f_{1,m})^2 m$. Using our coupled oscillator model and the experimentally determined particle-substrate contact stiffness $K_{N}$, we then solve for the effective interlayer contact stiffness $G_{e}$ of the measured twelve layer region (assuming that all interlayer effective contact stiffnesses are the same), by matching each of the measured modes shown in Figure~\ref{fgr:fig2}(b) to the corresponding eigenfrequency. Taking the average value of these effective interlayer stiffnesses, determined for each mode, we obtain a value of $G_{e}$ = $0.42$~kN/m. If only the fundamental mode is used, we obtain an effective interlayer thickness $G_{e}$ = $0.38$~kN/m, which is $8\%$ lower than the corresponding stiffness obtained by fitting all five measured modes. We note that these effective contact stiffnesses are similar to the stiffness found in Reference \cite{Fytas2012} for colloidal crystals composed of similar-sized PS spheres with FCC packing. Using these effective interlayer stiffnesses, we then recalculate and plot the resulting modes in Figure~\ref{fgr:fig2}(d), where the green open squares were calculated using the stiffness found by fitting only the fundamental mode and the closed circles using all five measured modes. In both cases, we see that the highest calculated mode deviates from the dispersion trend defined by the lower frequency modes due to the defect caused by the differing particle-substrate contact stiffness.\\

\begin{figure*}
 \begin{center}
 \includegraphics[width = 15cm]{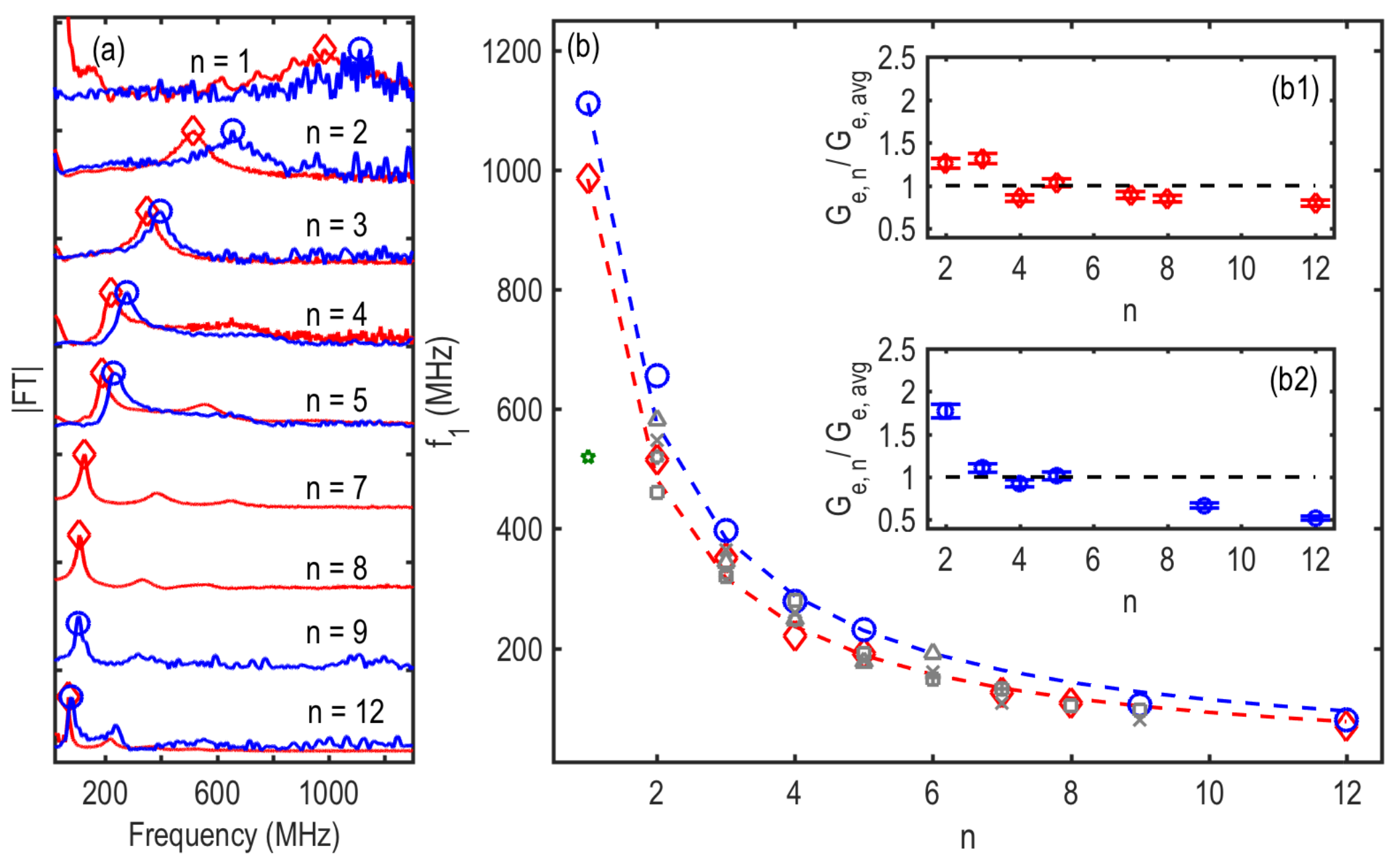}
 \end{center} 
 \caption{(a) Fourier Transform spectra of the time-resolved interferometric signals recorded from measurements on two colloidal crystal samples with thickness ranging from one to twelve layers. The Fourier Transform amplitude is plotted in linear scale, and offset for each layer thickness. (b) Frequencies of the fundamental mode plotted for colloidal crystals of varying number of layers ($n$) in the two samples from the peaks in the spectra in (a). The blue and red dashed lines indicate the frequencies of the fundamental eigenmode of a coupled oscillator system using the particle-substrate contact stiffness from the monolayer measurement ($K_{N}$) and the measured mean effective interlayer contact stiffness ($G_{e,avg}$). The gray markers represent frequencies of the fundamental modes on colloidal crystals fabricated with differing self-assembly parameters, but for which a monolayer resonance could not be measured. The green star marker represents the frequency measured on a monolayer that was pre-assembled at an air/water interface and subsequently transferred to a solid substrate. The inset highlights the variation of the effective interlayer contact stiffness for different layer thicknesses ($G_{e,n}$) in the two samples shown in (a). Red makers and lines correspond to Sample 1 and blue markers and lines to Sample 2 in all panels.}
 \label{fgr:fig3}
\end{figure*}

To similarly characterize the contact stiffnesses of colloidal crystals of different layer thicknesses, we perform systematic measurements on multiple samples with defined layer thicknesses, ranging from one to twelve layers, as is shown in \textbf{Figure~\ref{fgr:fig3}}. Each sample was also fabricated using different fabrication parameters, including particle concentration, solvent (deionized water or ethanol), temperature of the drying environment, and number of times the colloidal suspension was centrifuged and the supernatant discarded to remove impurities from the solution. A tabulated list of the fabrication parameters used for each sample is included in the Supporting Information. We note that, in contrast to the spectrum shown in Figure~\ref{fgr:fig2}(b), fewer peaks were observed for other measurement regions and samples. In addition to being a result of the higher number of averages ($10^6$) used for this measurement location, the presence of the additional peaks in Figure~\ref{fgr:fig2}(b) suggests reduced dissipation, which consequently may indicate lower disorder within that measurement area. This is further supported by an observed higher quality factor of the fundamental mode for this region ($Q \sim 12$), compared to other measured regions ($Q \sim 3-7$). The quality factor is obtained by fitting a Lorentzian distribution to the fundamental mode in the measured power spectra.\\

Figure~\ref{fgr:fig3}(a) shows the Fourier spectra measured for regions of different layer thicknesses across the two samples where we could obtain a measurement of the out-of-plane contact resonance in the monolayer region. The red curves in Figure~\ref{fgr:fig3}(a) correspond to the sample characterized in Figure~\ref{fgr:fig2}, which we refer to as ``Sample 1," and the blue to a different sample, which we refer to as ``Sample 2." Figure~\ref{fgr:fig3}(b) shows the frequency of the fundamental mode for all regions measured, as a function of the number of layers in that region. Each marker type corresponds to a different sample. The red and blue markers in Figure~\ref{fgr:fig3}(b) correspond to the spectra denoted by the same markers in Figure~\ref{fgr:fig3}(a). The gray markers denote samples for which a resonance was not detected in the monolayer region. For all samples we observe a similar, and expected, trend of decreasing fundamental mode frequency with increasing layer thickness. We suggest that this may be due to a combination of low quality factors observed for many of the monolayer regions, as can be seen by the broad peaks of the monolayer spectra in Figure~\ref{fgr:fig3}(a), and the possibility that the monolayers have frequencies above our detection bandwidth of $\sim 1$ GHz. As a point of comparison to the samples fabricated using the vertical deposition convective self-assembly technique and prior studies on monolayer contact resonances \cite{Boechler_PRL, EliasonMicrosphereStrip, HiraiwaThreeResonance, KhanolkarLambWaves2015}, we fabricated and measured a sample consisting entirely of a colloidal crystal monolayer using a modified Langmuir-Blodgett technique involving pre-assembly of the monolayer at an air/water interface and subsequent transfer to a solid substrate \cite{VogelLB}. The monolayer contact resonance on this air/water monolayer sample is found to be significantly lower than those measured for the samples fabricated using the vertical convective self-assembly technique, and is denoted by the green star marker in Figure~\ref{fgr:fig3}(b).\\

We repeat the previously described process to determine the effective interlayer contact stiffness for the different layer thicknesses tested on Sample 1 and 2. As the effective interlayer contact stiffness estimated using the fundamental and higher order modes was found to be reasonably close, and we do not have access to higher order modes for many of the regions tested, we fit only using the measured fundamental mode and the particle-substrate contact stiffness, determined from the monolayer region on the same sample, to find the effective interlayer contact stiffness for each thickness ($G_{e,n}$). The resulting effective interlayer contact stiffnesses, normalized by the average interlayer contact stiffness ($G_{e,avg}$) (across all layer thicknesses for the corresponding sample), are plotted in the insets Figure~\ref{fgr:fig3}(b1,b2) as a function of thickness. The half-width error bars for each marker in the insets represent the $8\%$ error in the effective interlayer stiffness calculated using the fundamental mode only or the first five eigenmodes in the coupled oscillator model. Each sample is denoted with the same marker type and color throughout Figure~\ref{fgr:fig3}.\\

The insets Figure~\ref{fgr:fig3}(b1,b2) highlight the variation of the effective interlayer stiffness in regions of differing thicknesses within the same sample. The effective interlayer stiffness measured on different layer thicknesses in Sample 1 shows little variation, with an average value of $0.44$~kN/m. On the other hand, we find that the effective interlayer stiffness decreases with increasing layer thickness in Sample 2, ranging from $1.2$~kN/m in the bilayer region to $0.35$~kN/m in the twelve-layer region of the sample. We speculate the decrease in stiffness could be due to increased disorder or a reduced density of impurities near the contact with increasing layer thicknesses, as has been previously suggested in Reference \cite{Ruello2017}. The presence of such impurities is to be expected as a by-product of colloidal synthesis processes. To verify the chemical nature of such impurities, we conducted IR spectroscopy on the material left over by separating the polystyrene particles from the colloidal dispersion, and then condensing and drying the supernatant (see Supporting Information). In addition to signals of polystyrene, the IR spectrum of this material showed additional peaks associated with co-monomer groups, which confirmed the presence of short-chained polymers in the colloidal dispersion. Indeed, spectroscopic and interfacial analyses of polymer colloidal dispersions reported previously in the literature \cite{Goodall1977}, have shown that impurities, consistent with what was found in our IR spectroscopic measurements, occur in emulsion polymerization processes, and consist of low molecular weight polymers that form during the process and remain water soluble. During the assembly process, these impurities will be co-deposited by capillary forces at the contact points between the colloidal particles and at the substrate/particle contact. Along these lines, we note that the colloidal suspension for Sample 2 was centrifuged three times, while the suspension for Sample 1 was only centrifuged once, as can be seen in the Supporting Information. The average interlayer stiffness calculated from measurements on all regions of Sample 2 is $0.68$~kN/m.\\

The red and blue dashed lines in Figure~\ref{fgr:fig3}(b) indicate the resulting fundamental eigenfrequencies of coupled oscillator systems with two to twelve masses connected to each other via contact springs having stiffness equal to the measured average interlayer stiffness $G_{e,avg}$ and to the substrate via a contact spring of stiffness equal to the measured particle-substrate contact stiffness $K_{N}$ for Sample 1 and 2, respectively. We find that the measured frequencies follow the trend predicted by the coupled oscillator model, including in the case of the samples where a monolayer resonance was not detected.\\

We find considerable disparity between the monolayer contact resonance frequency measurements for the two colloidal crystal samples ($\sim 1$~GHz) and that for the monolayer fabricated by the pre-assembly at the air/water interface ($520$~MHz). Table 1 lists the average effective interlayer stiffness and the particle-substrate contact stiffness calculated from the measured frequencies for the two multilayer samples and the air/water monolayer sample. Potential causes for this disparity, as well as comparisons with adhesive contact models will be discussed in the following.\\

We compare the average effective interlayer stiffness and the particle-substrate contact stiffness obtained from our measurements to estimates made using the Derjaguin-Muller-Toporov (DMT) adhesive elastic contact model \cite{DMT1}. The DMT adhesive elastic contact model is derived from the Hertz contact model \cite{HertzPaper}, with the addition of a static adhesive force. Based on the DMT model contact mechanics, the linearized normal contact stiffness around the equilibrium displacement is expressed as $K_{DMT} = \frac{3}{2} (2\pi w R_{e}^2 {E^{*}}^{2})^{1/3}$ \cite{Boechler_PRL}, where $w$ is the work of adhesion between the two surfaces in contact \cite{Israelachvili}, $R_{e}$ is the effective radius (equal to the radius of the sphere for the particle-substrate contact, and half the radius of the sphere for the particle-particle contact) and $E^{*} = [\frac{3}{4}((1 - \nu_{P}^{2})/E_{P} + (1 - \nu_{S}^{2})/E_{S})]^{-1}$ is the effective modulus of the contact. Using a work of adhesion between the polystyrene spheres and the silica-coated substrate obtained using the Lifshitz theory of van der Waals forces \cite{Israelachvili} of $w_{P-S}$ = $0.06$~J/$\text{m}^2$, and $\nu_{P}$ = $0.32$, $\nu_{S}$ = $0.17$, $E_{P}$ = $4.04$~GPa and $E_{S}$ = $73$~GPa as the Poisson's ratio and Young's modulus of the polystyrene sphere \cite{KhanolkarLambWaves2015} and the substrate \cite{GlassProperties}, respectively, we estimate the linearized normal contact stiffness between the particle and the substrate to be $K_{N, DMT}$ = $0.1$~kN/m. Similarly, we find $G_{e,DMT}$ = $0.1$~kN/m, using a work of adhesion between the spheres of $w_{P-P}$ = $0.06$~J/$\text{m}^2$ (see Supporting Information).\\

From Table 1, we observe that the particle-substrate and the effective interlayer stiffnesses predicted by the DMT model are much lower than all the corresponding measured contact stiffnesses. The discrepancy between measured contact stiffnesses and those predicted by the DMT model has been reported previously for silica \cite{Boechler_PRL, HiraiwaThreeResonance} as well as polystyrene microspheres \cite{EliasonMicrosphereStrip, KhanolkarLambWaves2015},  and could be explained by the uncertainty in the work of adhesion between contacting surfaces \cite{Israelachvili}, plastic deformation \cite{Wang2011}, or solid \cite{Tomas2007} or liquid \cite{MittalJaiswal, Johannsmann2006, LopezReview2016} material bridges around the contacts.\\

\begin{figure*}
 \includegraphics[width = 16cm]{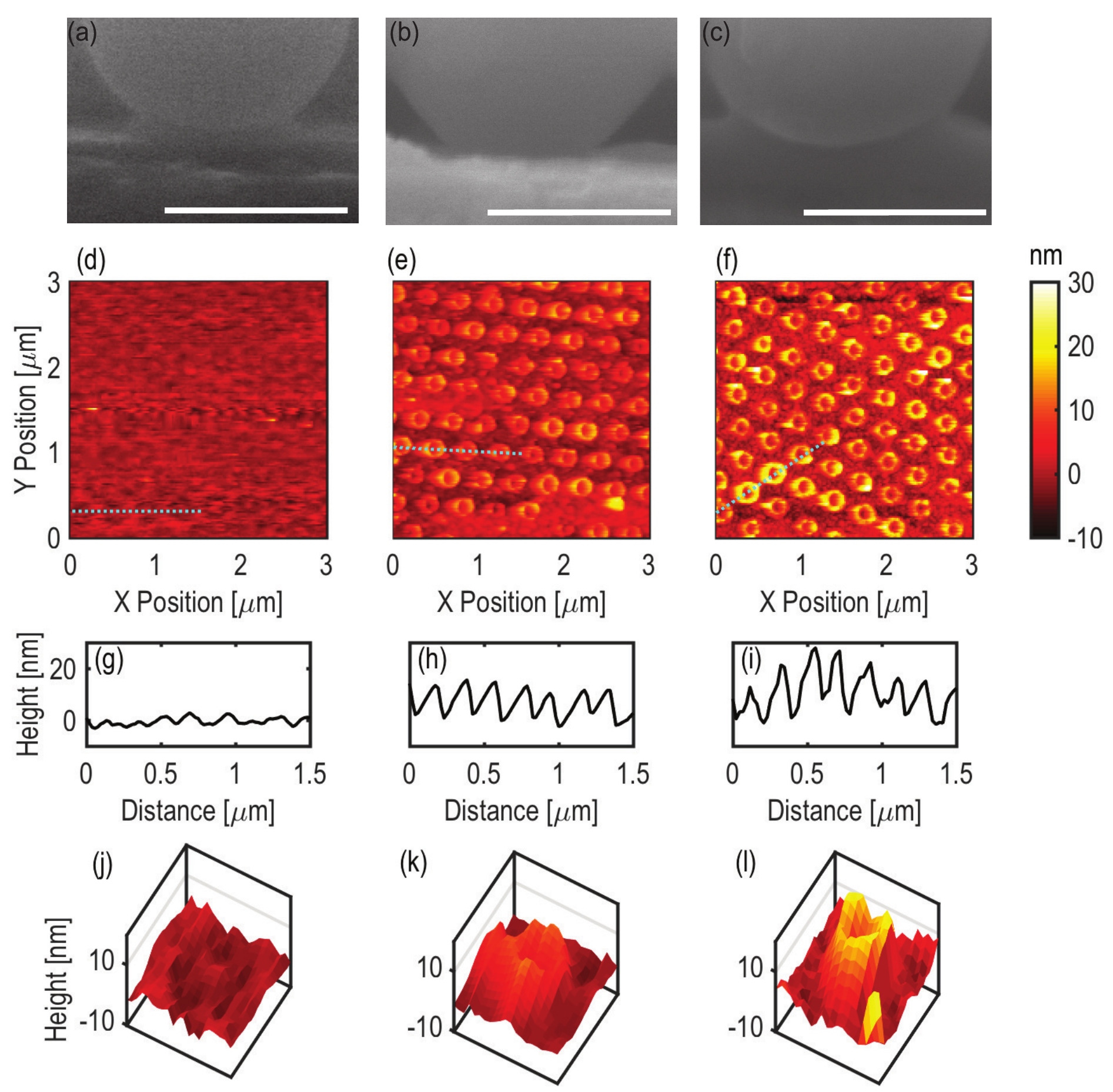}
 \caption{Scanning electron microscopy images of the monolayer regions in: (a) the air/water monolayer sample; (b) Sample 1; and (c) Sample 2. We note some minor lateral image distortion in the SEM image in panel (c). Scale bars represent 250 nm in all panels, however significant uncertainty should be assumed due to variations in the distance between the focal plane and the contact. (d) - (f) Tapping-mode Atomic Force Microscopy images of the substrate after removal of the spheres. (g) - (i) The surface topology of the substrate along the dashed line is shown in the corresponding panel directly above. (j) - (l) Isometric views of the AFM images of single `well'-like features on a $0.4$~$\mu$m x $0.4$~$\mu$m area of the substrate. All panels in the same column correspond to the same sample.}
 \label{fgr:fig4}
\end{figure*}

To gain insight into the source of the difference between the measured contact stiffnesses of the multilayer samples and the air/water monolayer sample, as well as that between the measurements and DMT model prediction, we analyze the contact surface by means of scanning electron microscopy. Representative side-view SEM images of the particle-substrate contacts in the air/water monolayer sample, and the monolayer regions of the two multilayer colloidal crystal samples investigated in Figure~\ref{fgr:fig3}(a) (i.e., Sample 1 and Sample 2) are shown in \textbf{Figure~\ref{fgr:fig4}}(a)-(c).  The SEM image of the air/water monolayer sample, shown in Figure~\ref{fgr:fig4}(a), indicates a large particle-substrate contact diameter (of $\sim 150$~nm), which is significantly higher than that predicted by the DMT model ($28$~nm). The large contact diameter in the air/water monolayer sample suggests that the spheres may have plastically deformed under the action of adhesive forces during the self-assembly process \cite{Wang2011}. The SEM image of the monolayer region of Sample 1, shown in Figure~\ref{fgr:fig4}(b), reveals a similar particle-substrate contact diameter as in the air/water monolayer sample. Although the particle-substrate contact diameters in the air/water monolayer sample and Sample 1 appear to be comparable, the particle-substrate contact stiffness measured in Sample 1 being over three times that measured in the air/water monolayer sample suggests that there may be other mechanisms responsible for the added stiffness of the contacts. Side-view SEM images of the particle-substrate contacts in the monolayer region of Sample 2, shown in Figure~\ref{fgr:fig4}(c), appear considerably different from the previous two cases. The spheres seem to be partially embedded in a solid matrix that forms a `well'-like structure around the particle. The cup-like structures appear to be formed from a thin film on the substrate. However, although the particle-substrate contacts in Samples 1 and 2 appear qualitatively different, the contact stiffnesses measured on these samples were comparable.\\

To further explore potential causes for the differing contact stiffnesses, we investigate the topography of the surface of the substrate after removal of the spheres by tapping-mode AFM. The spheres in the monolayer regions were removed by placing a piece of adhesive tape on the monolayer region, followed by gentle application of pressure on the surface of the tape. The spheres adhered to the tape when it was peeled off, leaving a region of blank substrate on the sample. The AFM images of the regions of the substrate where the monolayer was peeled off on the three samples are shown in Figure~\ref{fgr:fig4}(d)-(f). For the case of the air/water monolayer sample, the AFM image in Figure~\ref{fgr:fig4}(d) shows variations in the surface height within $\sim 3$~nm, which we attribute, in part, to the roughness of the silica film deposited on the substrate. However, the AFM image also reveals ring-like patterns of diameter $\sim 150$-$200$~nm at the sites previously occupied by the spheres, suggesting that these patterns could be due to residue on the surface. The corresponding AFM images of the multilayer colloidal crystal samples show a similar, but more pronounced, substrate surface topography. The AFM image of the substrate in Sample 1, shown in Figure~\ref{fgr:fig4}(e), shows very pronounced ring-like structures, $\sim 5$-$16$~nm in height, with hemispherical openings. Even more pronounced ring-like structures exceeding 20 nm in height are seen in the AFM image of Sample 2, which is illustrated in Figure~\ref{fgr:fig4}(f), and complements the relatively larger observed contact stiffness. The diameter of the hemispherical openings of the ring-like structures in Samples 1 and 2 is $\sim 160$-$200$~nm, similar to that of the ring-like structures observed on the air/water monolayer sample. Representative profiles of the surface height in the three samples are plotted in Figure~\ref{fgr:fig4}(g)-(h). Isometric close-up views of the ring-like structures revealed by the AFM images are shown in Figure~\ref{fgr:fig4}(j)-(l). A clear meniscus shape of solid material around a central depletion caused by the removed sphere is seen.\\

The SEM and AFM images of the regions around the particle-substrate contacts in the colloidal crystals compliment the contact stiffnesses obtained from the laser ultrasonic measurements, and provide possible explanations for the discrepancies in the particle-substrate contact stiffness between the air/water monolayer and multilayer samples, as well as discrepancies between our measurements and predictions made by the DMT contact model. For instance, the ratio of the contact stiffness between the air/water and Sample 2 monolayers was measured to be 3.3, as is shown in Table 1. We estimated that the difference in contact stiffness can be estimated by comparing the vertical particle deformation distance $d_1$ with the height of the added impurity $d_2$ material such that $K_2/K_1 = \sqrt{(1+d_2/d_1)}$. Using a particle deformation distance of $d_1 = 3$~nm, calculated from the measured contact resonance frequency of the monolayer, and assuming an impurity cup height of $d_2 = 20$~nm, we obtain a ratio of contact stiffness $K_2/K_1 = 2.7$, which is close to the measured difference. \\

\begin{figure*}
 \includegraphics[height = 5cm]{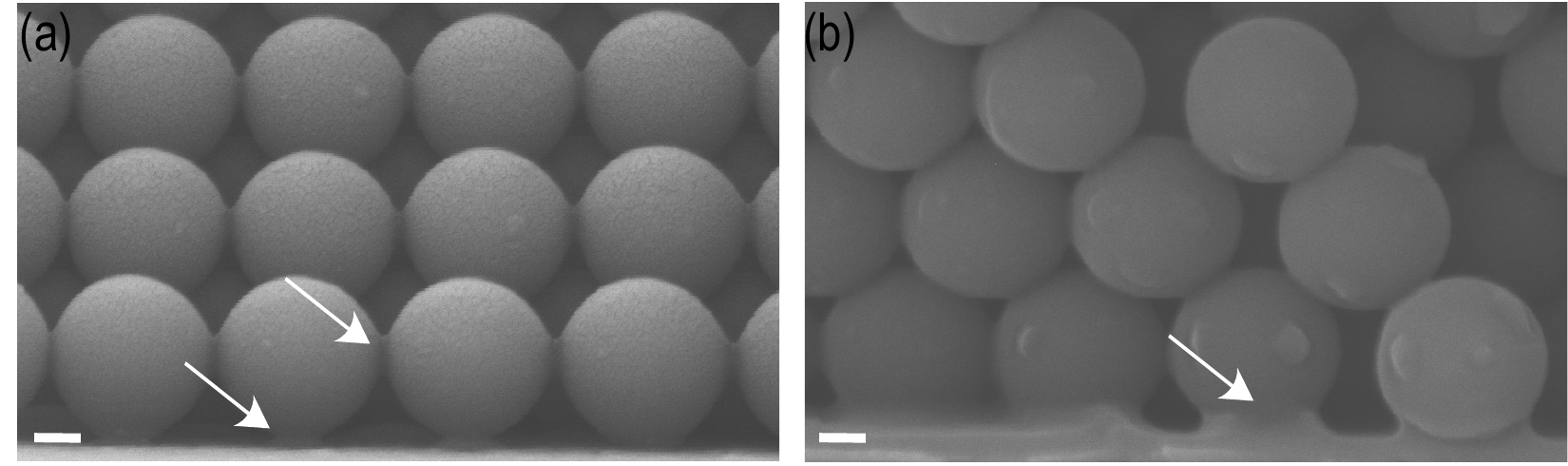}
 \caption{Scanning electron microscopy images illustrating particle-substrate and interparticle contacts in (a) a seven-layer-thick region of Sample 1 and (b) a six-layer-thick region of Sample 2. The scale bar is 100 nm in both panels. The arrows indicate representative material bridges observed between the contacts.}
 \label{fgr:fig5}
\end{figure*}

The previously described bridging can also be observed in the interparticle and particle-susbtrate contacts in the multilayer regions of the colloidal crystal samples. Representative views of the particle-substrate and interparticle contacts in the multilayer regions of Samples 1 and 2 are illustrated in \textbf{Figure~\ref{fgr:fig5}}(a) and (b), respectively. For both samples, a thin `bridging' material ($\sim 10$~nm in width) is seen in the region around the particle-substrate contact, as well in the interstitial space between neighboring spheres. As before, the `bridging' material is more evident in Sample 2 (shown in Figure~\ref{fgr:fig5}(b)), which can be seen as a continuous thin film along the substrate that forms cup-like structures around the particle-substrate contact. Similar to the case of the monolayer regions, the material is also deposited at the contact points between neighboring spheres. We attribute these solid bridges to the presence of solid impurities within the colloidal suspension.\\

The differences between the various samples in this study, as seen from measurements of the contact stiffnesses via laser ultrasonic characterization as well as in the SEM and AFM images of the contacts, can be attributed to the different nature of the self-assembly fabrication process. In the modified Langmuir-Blodgett technique, a floating monolayer of spheres is pre-assembled at an air/water interface, which is then transferred onto a substrate \cite{VogelLB}. Solid impurities in the colloidal suspension, such as water-soluble polystyrene oligomers, other reaction side-products from the emulsion polymerization process \cite{VogelSynthesis}, or polymer impurities that may leach from the vials containing the colloidal solution, are much more diluted as the water subphase provides an enormous reservoir. This may explain the smaller ring-like structures seen in the AFM images of the air/water monolayer sample, Figure~\ref{fgr:fig4}(d). On the other hand, the `bridging' material seen between the sphere contacts in Samples 1 and 2 may be formed by solid impurities occupying interstitial sites in the crystal lattice. Unlike the case of the air/water pre-assembly technique, it is possible that the solid impurities in the vertical deposition technique are pulled towards spheres that have already deposited on the substrate by convective forces. Subsequently, these impurities are concentrated in liquid capillary bridges that form in between the spheres and the substrate, where they finally solidify upon drying. As a result, the purity of the assembly suspension (i.e., the amount of impurities) will determine the quality of the colloidal crystal sample, which in turn would be expected to affect the contact properties, wherein cleaner suspensions should give less solid bridges and suspensions with more solid impurities should result in colloidal crystals with more solid bridges. As suggested earlier, multiple cycles of centrifuging the colloidal suspension and removing the supernatant prior to self-assembly deposition can alter the quantity of impurities in the colloidal suspension. The use of repeated centrifuge purification steps in the preparation of Sample 1 readily explains the slightly higher particle-substrate contact stiffness in Sample 2 over Sample 1, and the thicker film on the surface of the substrate. In addition to affecting the average interlayer contact stiffness of the multilayer, an increased density of impurities at the lower layers may also contribute to the observed decrease in stiffness with increasing layer thickness shown in Figure~\ref{fgr:fig3}(b2). Our measurements, therefore, open avenues to potentially assess the packing and contact quality in three-dimensional colloidal crystals.\\

For further insights into the mechanical wave propagation properties in our colloidal crystals, we estimate the long wavelength longitudinal sound speed $v$ in the colloidal crystal samples from the measured average interlayer contact stiffnesses using the relation $v = \sqrt{\frac{G_{e,avg}}{m}} D^{*}$, where $D^{*} = \frac{\sqrt{6}}{3} D$ is the unit cell spacing, and find $v_{1} = 1166$~m/s and $v_{2} = 1445$~m/s for Samples 1 and 2, respectively. Our estimated values are about half the longitudinal sound speed in bulk polystyrene, $v_{PS} = 2350$~m/s \cite{KhanolkarLambWaves2015}, and about two times higher than the sound speed calculated using the interlayer contact stiffness predicted by the DMT model, $v_{DMT} = 617$~m/s. The long wavelength longitudinal sound speed calculated from the average interlayer contact stiffness is consistent with the sound speed estimated from the observed contact area. This suggests that the contact stiffness can be reasonably estimated from the contact area, even when the force of adhesion is unknown. We also calculate the long wavelength sound speed obtained from the observed contact diameter, assuming DMT contact mechanics. Using a contact diameter of $150$~nm yields a sound speed of $1582$~m/s, which is in reasonable agreement with the sound speed calculated from the measured average interlayer stiffness. At the macroscale, an uncompressed granular crystal is often described as a sonic vacuum (i.e., zero sound speed) wherein the particle-particle interactions are purely nonlinear \cite{PhysicsTodayGranular}. As this system is further compressed, it can be considered to linearize, with a finite low-amplitude sound speed. Compared to sound speeds as low as $\sim 200$~m/s measured in slightly compressed macroscale granular crystals \cite{Coste}, the relatively high speeds in our samples suggest a highly linearized system. Strategies to increase the coupling between macroscale spheres in contact and therefore modify their dynamic response have been studied previously, for instance by welding a finite chain of millimeter-sized steel spheres \cite{HladkyHennion_WeldedSpheres}, where the contact diameter was $\sim 30\%$ of the sphere diameter. Controlling the size of the material bridges between the spheres may therefore be used as an analogous mechanism to tune the coupling between particles at the micro- and nanoscale.\\

Finally, we compare the quality factors of the peaks of the measured eigenfrequencies to those estimated from the acoustic impedance mismatch between the colloidal crystal and the glass substrate. We estimate the acoustic impedance of the colloidal crystal to be $Z_{1} = \rho_{c} v_{1} = 0.92$~MPa.s/m, where $v_{1}$ is the longitudinal sound speed calculated previously for Sample 1 using the average interlayer contact stiffness and effective density of the colloidal crystal of $\rho_{c} = \frac{6\phi m}{\pi D^3} = 784$~kg/m$^{3}$, which is calculated using a solid volume fraction $\phi = 0.74$ assuming HCP packing. Similarly, we calculate the acoustic impedance of the glass substrate to be $Z_{s} = \rho_s v_s = 14.2$~MPa.s/m, where $\rho_s=2500$ kg/m$^3$ is the density of the substrate and $v_s=5697$ m/s is its sound speed \cite{Substrateproperties}. The resulting amplitude reflection coefficient $r$ at the colloidal crystal/ substrate interface is calculated to be $r = \frac{Z_{s}-Z_{1}}{Z_{s}+Z_{1}} = 0.88$ \cite{Auld}. We estimate the expected quality factor of the fundamental mode, accounting for radiation from the colloidal layer, using the relation $Q_{i} = \frac{\pi(2i-1)}{1-r^2}$. For the fundamental mode ($i=1$), we estimate a quality factor of $Q_{1} = 14$, which is slightly larger than the observed quality factor for the fundamental mode of the $12$ layer thick region of Sample 1, and respectively larger than the other measured fundamental modes. Furthermore, for the 12 layer thick region of Sample 1, the measured quality factors of the higher-order peaks vary between $\sim 7 - 10$, which is in contrast to the increase in quality factor with mode number that is expected for damping stemming from radiation into the substrate. The observed quality factors for the higher-order, shorter-wavelength modes thus suggest additional susceptibility to material or disorder-related loss mechanisms, including mechanisms arising from disorder in the contact stiffnesses.\\

\textbf{3.	Conclusion}\\
 %%=========================
%% ==Conslusion===
%%=========================

In summary, we have studied the contact-based longitudinal eigenvibrations of ordered, three-dimensional colloidal crystals adhered to solid substrates. We identify regions of uniform layer thickness in the colloidal crystal by their structural color. This enables us to measure the longitudinal eigenvibrations of the colloidal crystal as a function of layer thickness. Using non-contact laser ultrasonic measurements, along with a coupled oscillator model, we extract the particle-substrate and effective interlayer contact stiffness in the samples, and in one case find the effective contact stiffness between each layer to decrease with increasing number of layers. The laser ultrasonic measurements, supplemented by SEM and AFM images, show that nanometric-sized bridges around the contacts can drastically affect the contact stiffnesses. In the future, we expect that the rational control of contact stiffness via deposition of solid bridges can be systematically exploited to tailor the acoustic properties of self-assembled structures, which may in turn lead to a new class of tunable `micro- and nanoscale granular crystals.' This study furthers the understanding of the dynamics of self-assembled micro- and nanoparticulate structures, which may have potential for ultrasonic wave tailoring applications analogous to macroscale granular crystals.\\

%%%%%%%%%%%%%%%%%%%%%%%%%%%%%%%%%%%%%%%%%%%%%%%%%%%%%%%%%%%%%%%%%%%%% %%%%%%%%

\textbf{4.	Experimental Section}\\
 %%=========================
%% ==Methods===
%%=========================

\textbf{Colloidal Crystal Fabrication}: We use monodisperse polystyrene spheres of diameter $D$ = $390$~nm that form a colloidal crystal with high order. The colloidal crystal is assembled on a substrate that consists of a $1.5$ mm glass microscope slide, which is coated with a 100 nm thick layer of aluminum to absorb the optical pump energy and then a $20$ nm thick silica layer to facilitate the self-assembly process. The polystyrene spheres are synthesized using a surfactant-free emulsion polymerization process \cite{VogelSynthesis}. The crystals are fabricated using a vertical convective self-assembly technique \cite{JiangSelfAssembly, MeijerSelfAssembly}, using varied solvents and purification strategies, as is detailed in the Supporting Information. A $20$~mL capacity glass scintillation vial is filled with $10$~mL of the colloidal suspension. The substrate is then held vertically while being immersed in the suspension.  The substrate and suspension are then left to dry in an oven or in ambient laboratory conditions (see Supporting Information).  After complete drying, colloidal crystals with millimeter areas of defined and uniform thicknesses are obtained.\\

 \textbf{Laser Ultrasonic Characterization}: Optical pump pulses ($450$ ps pulse duration, $532$~nm wavelength, $7$~$\mu$J pulse energy, and $1$~kHz repetition rate) incident through the glass slide are focused onto the aluminum film. The pump pulse is focused to an elliptical spot ($436$~$\mu$m x $76$~$\mu$m at $1/e^2$ intensity level) or a $200$~$\mu$m diameter circular spot, depending on the size of the layer being characterized (see Supporting Information). The vibrations are detected with a phase-mask-based interferometer \cite{GlorieuxInterferometer2004} in which a single continuous wave (CW) laser beam ($514$~nm wavelength and  $52$~mW average power) is incident on a phase-mask and split into $+/-1$ diffraction orders to produce probe and reference beams focused to ~$40$~$\mu$m diameter spots. The probe is focused through the colloidal film onto the aluminum surface, while the reference beam is incident directly on the aluminum surface of a blank region of the substrate. The separation between the probe and reference beams is $4$~mm. Upon reflection from the sample, the probe and reference are recrossed onto the phase mask, and recombined interferometrically onto an amplified silicon photodetector where the signal is digitized and recorded using an oscilloscope, and averaged over $10^4$ - $10^6$ pump pulses. \\

%%%%%%%%%%%%%%%%%%%%%%%%%%%%%%%%%%%%%%%%%%%%%%%%%%%%%%%%%%%%%%%%%%%%% %%%%%%%%

\textbf{Supporting information} \\
Supporting Information is available from the Wiley Online Library or from the author.\\

%%%%%%%%%%%%%%%%%%%%%%%%%%%%%%%%%%%%%%%%%%%%%%%%%%%%%%%%%%%%%%%%%%%%% %%%%%%%%

\textbf{Acknowledgements} \\
We acknowledge support from the U.S. Army Research Office (grant no. W911NF-15-1-0030) and the U.S. National Science Foundation (grant no. CMMI-1333858). N.V. acknowledges funding by the Deutsche Forschungsgemeinschaft (DFG) through the Cluster of Excellence ``Engineering of Advanced Materials'' (EXC 315) and support by the Interdisciplinary Center for Functional Particle System (FPS) at FAU Erlangen. A.M. acknowledges support from the U.S. Department of Energy (grant no. DE-FG02-00ER15087). We acknowledge contributions to sample fabrication from Sergei Romanov. We also thank A. Emery and V. Holmberg for useful discussions, and M. Glaz for the AFM measurements. Part of this work was conducted at the Molecular Analysis Facility, a National Nanotechnology Coordinated Infrastructure site at the University of Washington which is supported in part by the National Science Foundation (grant ECC-1542101), the University of Washington, the Molecular Engineering \& Sciences Institute, the Clean Energy Institute, and the National Institutes of Health. M.A.G and A.K. contributed equally to this work.\\

%%%%%%%%%%%%%%%%%%%%%%%%%%%%%%%%%%%%%%%%%%%%%%%%%%%%%%%%%%%%%%%%%%%%% %%%%%%%%

%===================
%== bibliography ===
%===================
%%%REFERENCES%%%

%%%%%%%%%%%%%%%%%%%%%%%%%%%%%%%%%%%%%%%%%%%%%%%%%%%%%%%%%%%%%%%%%%%%% %%%%%%%%

%===================
%== Figures and Tables ===
%===================
\newpage

\clearpage

\clearpage

\clearpage

\clearpage

\clearpage

\begin{table}[h]
\label{table:contact_stiffnesses}
\begin{center}
\begin{tabular}{ |M{3.0cm}|M{1.5cm}|M{1.5cm}|} 
 \hline
  & $K_{N}$ (kN/m) & $G_{e,avg}$ (kN/m) \\ 
 \hline
 Sample 1 & $1.3$  & $0.4$ \\ 
 Sample 2 & $1.6$ & $0.7$  \\ 
 Air/water monolayer & $0.4$ & - \\ 
 \hline
 DMT Model & $0.1$ & $0.1$ \\ 
 \hline
\end{tabular}
\vspace{10mm}
\end{center}
\caption{Measured and predicted (using the Derjaguin-Muller-Toporov, or ``DMT'' adhesive elastic contact model) particle-substrate and average interlayer contact stiffness. The DMT model assumes $w_{P-S}$ = $0.06$~J/$\text{m}^2$ and $w_{P-P}$ = $0.06$~J/$\text{m}^2$.}
\end{table}

\clearpage

%%%%%%%%%%%%%%%%%%%%%%%%%%%%%%%%%%%%%%%%%%%%%%%%%%%%%%%%%%%%%%%%%%%%% %%%%%%%%

%===================
%== Supporting Info===
%===================

\textbf{Supporting Information}\\

\noindent{\textbf{Longitudinal Eigenvibration of Multilayer Colloidal Crystals and the Effect of Nanoscale Contact Bridges}} \\

\noindent{\textit{Maroun Abi Ghanem, Amey Khanolkar, Samuel P. Wallen, Mary Helwig, Morgan Hiraiwa, Alexei A. Maznev, Nicolas Vogel and Nicholas Boechler$^{*}$}}\\
\\

\textbf{1.	Sample Fabrication}:
The aluminum-coated soda lime glass substrates were cleaned by soaking in isopropanol and acetone for ten minutes each, and then rinsed with deionized (DI) water and dried under nitrogen flow. Following this, the substrates were coated with a $20$-nm-thick silica layer via chemical vapor deposition. The colloidal suspensions for these samples were purified by centrifuging the suspension at $4000$ rpm, discarding the supernatant and redispersing the particles in DI water or ethanol.  This purification procedure was repeated three times for all samples, except in the case of Sample 2, where the colloidal suspension was centrifuged once.  Details of the self-assembly parameters and the layer thicknesses characterized for the multilayer samples in this study are listed in Table S1.  The samples with the microsphere suspension in water were left to dry in an oven at $75$\degree~C, while the samples immersed in the ethanol suspension were placed under a plastic container to avoid external airflows and left to dry under ambient laboratory conditions.

\setcounter{table}{0}
\begin{table}[H]
\renewcommand{\thetable}{S\arabic{table}}
\label{table:samples}
\caption{Details of sample fabrication parameters for the multilayer samples.}
\begin{center}
\begin{tabular}{ |c|P{2.1cm}|P{1.5cm}|P{2cm}|c|P{3.5cm}|P{2cm}|} 
 \hline
 Sample & Concentration (\% v/v) & Solvent & Drying Environment & Centrifuged & Layer Thicknesses Characterized & Monolayer Resonance Detected\\ 
 \hline
 1 & $0.2$ & Ethanol & Ambient & 3x & 1, 2, 3, 4, 5, 7, 8, 12 & Yes \\ 
 2 & $0.01$ & DI Water & Oven, $75$\degree~C & 1x & 1, 2, 3, 4, 5, 9, 12 & Yes \\ 
 \hline
 3 & $1$\ & Ethanol & Ambient & 3x & 6, 7 & No \\ 
 4 & $0.5$ & Ethanol & Ambient & 3x & 2, 3, 4, 5, 6, 7, 8, 9 & No \\ 
 5 & $0.3$ & Ethanol & Ambient & 3x & 2, 3, 4, 5, 6, 7, 9 & No \\ 
 6 & $0.01$ & DI Water & Oven, $75$\degree~C & 3x & 2, 3, 4 & No \\ 
 7 & $0.01$ & DI Water & Oven, $75$\degree~C & 3x & 2, 3, 4, 5, 6 & No \\ 
 \hline
\end{tabular}
\end{center}
\end{table}

\textbf{2.	Laser Ultrasonic Setup Details}:
For the measurements on Sample 2, the pump beam was focused to an elliptical spot ($436$~$\mu$m x $76$~$\mu$m at the $1/e^2$ intensity level), whereas for all the other samples, the pump beam was focused to a $200$~$\mu$m diameter circular spot.  The interferometric signals were averaged over $10^4$ pump pulses for all samples, except in the case of the measurements on the monolayer, five-, seven-, eight- and twelve-layer regions of Sample 1, which were averaged over $10^6$ pump pulses to achieve an improved signal-to-noise ratio.\\

\textbf{3.	Signal Processing Procedure}:
Time-resolved signals were recorded at a positive and negative phase setting by varying the optical path difference between the probe and the reference beams via a rotating fused silica window placed in the path of the probe beam. The net signal was obtained by subtracting the averaged waveforms collected at positive and negative phase settings.  A segment of $0.75$ $\mu$s of the signal starting from the sharp initial rise (corresponding to the arrival of the pump pulse) was used for further signal processing.  This segment of the signal was zero-padded after the oscillations completely decayed below the noise floor, differentiated with respect to time to remove the thermal decay component from the signal, and normalized with respect to its maximum amplitude. A Fast Fourier Transform (FFT) was then applied to the normalized signal. \\

\textbf{4.	Work of adhesion at the sphere-substrate and sphere-sphere interfaces}:
We estimate the work of adhesion between identical polystyrene microspheres $w_{P-P}$ and that between the polystyrene microspheres and the silica-coated substrate $w_{P-S}$ in terms of the Hamaker constant $A$, using $w=\frac{A}{12\pi D_0^2}$ ($D_0=0.165$~nm is a standard value used for the interfacial cutoff separation distance for a variety of media \cite{Israelachvili}). This coefficient takes into account the van der Waals forces between the two surfaces in contact. The expression for the Hamaker constant $A_{131}$ for two polystyrene surfaces (denoted as medium $1$) interacting across medium $3$ (air) is: 
\begin{equation}
A_{131}=\frac{3kT}{4}(\frac{\epsilon_1-\epsilon_3}{\epsilon_1+\epsilon_3})^2+\frac{3h\nu_{e1}}{16\sqrt{2}}\frac{(n_1^2-n_3^2)^2}{(n_1^2+n_3^2)^{3/2}}
\label{Hamaker}
\end{equation}
where $T=293$ K is room temperature, $\nu_{e1}=2.3\cdot10^{15}$ Hz is the main electronic absorption frequency of polystyrene in the UV, $k$ is Boltzmann's constant, $h$ is Planck's constant, $\epsilon_1=2.55$~F/m and $\epsilon_3=1.0$~F/m are the permittivities of polystyrene and air, respectively, and $n_1=1.56$ and $n_3=1.00$ are the refractive indices of polystyrene and air, respectively. All of the preceding constants are tabulated in Reference\cite{Israelachvili}. Using Eq.~\ref{Hamaker}, we calculate $A_{131}=6.53\cdot10^{-20}$ J. This gives a work of adhesion $w_{P-P}=0.06$ J/$\text{m}^2$ between the polystyrene microspheres.

Similarly, we use Eq.~\ref{Hamaker} to calculate the Hamaker constant for two silica surfaces (denoted as medium $2$) interacting in air (medium $3$), with the corresponding material properties of silica: permittivity $\epsilon_2=3.8$~F/m, refractive index $n_2=1.45$, and electronic absorption frequency in the UV $\nu_{e2}=3.2\cdot10^{15}$ Hz \cite{Israelachvili}. We obtain $A_{232}=6.36\cdot10^{-20}$ J. We use a combining relation to obtain an approximate value for the Hamaker constant between polystyrene and silica, $A_{132} \approx \sqrt{A_{131}A_{232}}$, such that $A_{132} \approx 6.44\cdot10^{-20}$ J. The corresponding work of adhesion between polystyrene and silica is $w_{P-S}=0.06$ J/$\text{m}^2$.  \\

\textbf{5.	Calculation of the effective interlayer contact stiffness using the DMT Model}:
The normal contact stiffness between two colloidal particles $G_{N,DMT}$ is derived from the linearized force-displacement curve prescribed by the DMT model \cite{DMT1}, and is expressed as, $G_{N,DMT} = \frac{3}{2} ( 2 \pi w_{P-P} R_{e}^2 {E^{*}_{P-P}}^2 )^{1/3}$, where $w_{P-P}$ is the work of adhesion between the two polystyrene surfaces, $R_{e}$ is the effective radius (equal to half the radius of the particle), and $E^{*}_{P-P}$ is the effective modulus of the contact and is defined in terms of the Young's modulus $E_{P}$ and the Poisson's ratio $\nu_{P}$ of the particle, $E^{*}_{P-P} = \frac{2}{3} \frac{E_{P}}{1-{\nu_{P}}^2}$. The effective interlayer normal contact stiffness $G_e,DMT$ is then derived by accounting for the contributions of the three normal $G_{N,DMT}$ and transverse $G_{S}$ contact springs each in an HCP unit cell along the out-of-plane direction to be $G_{e,DMT} = G_{N,DMT} (2 + \nu^*)$, where $\nu^* = \frac{G_{S}}{G_{N,DMT}} = 2 \frac{1-\nu_{P}}{2-\nu_{P}}$ is the ratio of the interparticle shear and normal contact stiffnesses \cite{Mindlin,Merkel}. \\

\textbf{6.	IR spectroscopy of colloidal particles and impurities}:
We used IR spectroscopy to obtain chemical information regarding the impurities by separating the polystyrene particles from the colloidal dispersion, and then condensating and drying the supernatant to analyze the impurities. The resulting IR spectrum of both components is presented in Figure~\ref{fgr:figS1}. 

\setcounter{figure}{0}

\begin{figure*}
\renewcommand{\thefigure}{S\arabic{figure}}
 \includegraphics[height = 7cm]{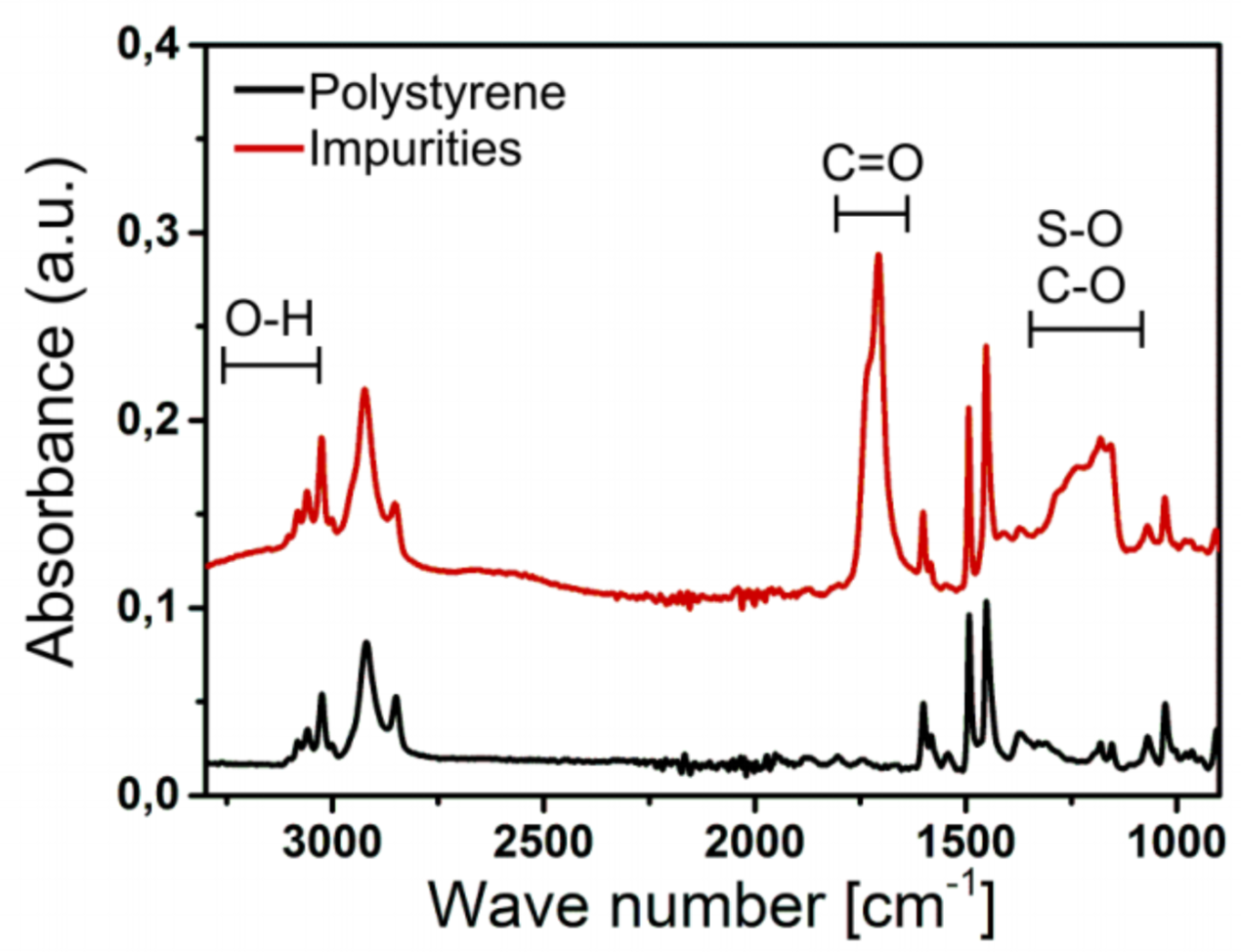}
 \caption{IR spectrum of polystyrene and the supernatant of the colloidal dispersion.}
 \label{fgr:figS1}
\end{figure*}

We see that the IR spectrum of the colloidal solution shown in Figure~\ref{fgr:figS1} contains all the signals of polystyrene, but additional OH and carbonyl groups (both can be traced to acrylic acid as the co-monomer), and S-O bonds that can be assigned to the sulfate groups from the initiator (potassium peroxodisulfate).


\begin{thebibliography}{99}
\bibitem{FootnoteColloid}
We note that such ``dry'' colloidal crystals are not strictly colloids, even though they are typically prepared from a colloid.
\end{thebibliography}
\end{document}